\documentclass[aps,pre,twocolumn,superscriptaddress,floatfix,nofootinbib]{revtex4-1}
\usepackage{amsmath,amssymb,color,comment,ulem}
\usepackage{graphicx}
\usepackage[english]{babel}
\usepackage[bookmarks=true,colorlinks,linkcolor=blue,urlcolor=blue,citecolor=blue]{hyperref}

\usepackage{bm}
\usepackage{braket}

\newcommand{\unity}{\openone}

\newcommand{\rmi}{{\rm i}}
\newcommand{\Fig}[1]{Fig.~\ref{#1}}
\newcommand{\Sec}[1]{Sec.~\ref{#1}}

\newcommand{\hczII}{\ensuremath{h^{\rm II}_{\rm c,z}}}

\def\figpath{.}

\begin{document}
% \title{Probing the anomalous dynamical phase in long-range quantum spin chains through Fisher-zero lines}
\title{Probing the anomalous dynamical phase in long-range quantum spin chains through Fisher-zero lines}
\author{Valentin Zauner-Stauber}
\affiliation{Vienna Center for Quantum Technology, University of Vienna, Boltzmanngasse 5, 1090 Wien, Austria}
\author{Jad C. Halimeh}
\affiliation{Physics Department and Arnold Sommerfeld Center for Theoretical Physics, Ludwig-Maximilians-Universit\"at M\"unchen, D-80333 M\"unchen, Germany}
\affiliation{Max Planck Institute for the Physics of Complex Systems, 01187 Dresden, Germany}

\date{\today}

\begin{abstract}
Using the framework of infinite Matrix Product States, the existence of an \textit{anomalous} dynamical phase for the transverse-field Ising chain with sufficiently long-range interactions was first reported in [J.~C.~Halimeh and V.~Zauner-Stauber, arXiv:1610:02019], where it was shown that \textit{anomalous} cusps arise in the Loschmidt-echo return rate for sufficiently small quenches within the ferromagnetic phase. In this work we further probe the nature of the anomalous phase through calculating the corresponding Fisher-zero lines in the complex time plane. We find that these Fisher-zero lines exhibit a qualitative difference in their behavior, where, unlike in the case of the regular phase, some of them terminate before intersecting the imaginary axis, indicating the existence of smooth peaks in the return rate preceding the cusps. Additionally, we discuss in detail the infinite Matrix Product State time-evolution method used to calculate Fisher zeros and the Loschmidt-echo return rate using the Matrix Product State transfer matrix. Our work sheds further light on the nature of the anomalous phase in the long-range transverse-field Ising chain, while the numerical treatment presented can be applied to more general quantum spin chains.
\end{abstract}
\maketitle

\section{Introduction}
Equilibrium phase transitions are a hallmark of statistical mechanics and condensed-matter physics. They are well-understood textbook subjects that have been treated through various theories. From Landau's theory of symmetry breaking \cite{Cardy,Sachdev} that presents an intuitive field-theoretical approach built upon the symmetries of the model under consideration to Wilson's renormalization group \cite{Fisher1974} that utilizes scale invariance, the theoretical description of equilibrium phase transitions has been validated in numerous experiments, not the least in quantum many-body systems. Nevertheless, the field of \textit{dynamical} phase transitions, particularly in quantum many-body systems, is still in its infancy, and much is left unanswered. 

One of the most interesting notions of dynamical criticality in quantum many-body systems in the thermodynamic limit is that manifested in non-analyticities in the Loschmidt-echo return rate \cite{Heyl2013,Heyl2015} of a quenched system. We have termed this type of dynamical phase transition as DPT-II \cite{Halimeh2016b}, in contrast to dynamical phase transitions defined in terms of dynamical order parameters (DPT-I). One of the paradigmatic models of statistical mechanics, the one-dimensional transverse-field Ising model (TFIM), has been the bulwark of investigations of the DPT-II in its nearest-neighbor (NN-TFIM) \cite{Heyl2013,Heyl2015} and long-range interacting (LR-TFIM) \cite{Zunkovic2016,Zunkovic2016b,Halimeh2016b,Halimeh2017} variants. For short-range interactions the Loschmidt-echo return rate displays cusps for quenches across a dynamical critical point (\textit{regular} phase) and no cusps otherwise (\textit{trivial} phase), where for nearest-neighbor interactions in particular the equilibrium and dynamical critical points coincide \cite{Heyl2013,Zunkovic2016b,Halimeh2016b}. With sufficiently long-range interactions, however, the Loschmidt-echo return rate shows cusps also for quenches within the equilibrium ordered phase that do not cross the dynamical critical point, with these cusps being of a different nature \cite{Halimeh2016b,Halimeh2017}.

In particular, considering in the case of long-range interactions only quenches from an ordered initial state, it has been shown in Refs.~\onlinecite{Halimeh2016b,Halimeh2017} that the Loschmidt-echo return rate displays \textit{anomalous} cusps for quenches below the dynamical critical point (which in this case does not coincide with the equilibrium critical point), whereas for quenches above it, the traditional \textit{regular} cusps familiar from the NN-TFIM case are present. Besides other qualities distinguishing the anomalous cusps from their regular counterparts, the defining feature of the anomalous phase is that the Loschmidt-echo return rate displays cusps only after its first minimum. Especially for small quenches these cusps typically only appear at rather late times after a number of preceding smooth peaks. Moreover, at least in the limit $\alpha=0$ the anomalous (regular) phase coincides with a finite (zero) long-time average of the $\mathbb{Z}_2$ order parameter \cite{Halimeh2017}.

In this paper we seek to shed more light on the qualitative difference between the anomalous and regular cusps, in particular in terms of Fisher zeros, as computed in the framework of infinite Matrix Product States (iMPS). Much like the Lee-Yang analysis of equilibrium phase transitions in the complex external-magnetic-field plane \cite{Yang1952,Lee1952} where a phase transition exists only if Lee-Yang zeros (or more specifically, the Lee-Yang \textit{singularity edges} \cite{Kortman1971,Fisher1978}) cross the positive real axis in the thermodynamic limit, the partition function is an entire function in the complex temperature plane \cite{Fisher1965} where Fisher zeros cross the real axis at a phase transition in the thermodynamic limit. Fisher zeros in the context of the DPT-II were first discussed in Ref.~\onlinecite{Heyl2013}, where it was shown that Fisher zeros cut the time axis, \textit{i.e.}~a dynamical phase transition exists, when the quench is across the equilibrium critical point in the case of the NN-TFIM. However, the equilibrium critical point is not always the dynamical critical point \cite{Andraschko2014,Vajna2014}, and the latter departs further from the former with increased range of interactions in the LR-TFIM \cite{Zunkovic2016b,Halimeh2016b,Halimeh2017}.

The paper is organized as follows: In Sec.~\ref{sec:IMPS} we discuss in some detail the workings of the iMPS method and its use in calculating our results. In Sec.~\ref{sec:FZ} we present the Fisher-zero data, and discuss the qualitative difference between the anomalous and regular cases. In Sec.~\ref{sec:appendix} we discuss the existence of \textit{double cusps} in the anomalous phase and we illustrate the crossover between the anomalous and regular phases around the dynamical critical point, both in terms of the level crossings of the rate-function branches of the Matrix Product State (MPS) transfer matrix. We conclude in Sec.~\ref{sec:conclusion}.

\section{Infinite Matrix Product State technique}
\label{sec:IMPS}
% \subsection{Motivation}
As shown in the seminal works \cite{Heyl2013,Heyl2015} for the NN-TFIM, a DPT-II manifests itself as non-analyticities in the form of cusps in the Loschmidt-echo return rate. Due to finite-size effects, these cusps will be smoothened out in finite systems, thus indicating no non-analyticity (and thus no criticality) in the Loschmidt-echo return rate. One can in principle employ conventional MPS time-evolution techniques for finite-size systems \cite{TEBD_supp,tDMRG1_supp,tDMRG2_supp} and use finite-size scaling in order to surmise a cusp in the thermodynamic limit, but such a procedure might be error-prone, cumbersome, and requires unrealistic numerical resources. It is therefore desirable to perform simulations directly in the thermodynamic limit, which can be done very efficiently using iMPS \cite{MPS1_supp,MPS2_supp, Uli_supp} without any approximations in addition to using Matrix Product State representations. One can then directly see sharp cusps in the Loschmidt-echo return rate, which are due to level crossings in the eigenvalues of the MPS transfer matrix (see below). This constitutes an additional advantage over a finite-size-scaling approach, where it can be very unclear if a maximum in the return rate will develop into a non-analytic cusp in the limit of infinite system size. In the case of iMPS, due to translation invariance, the presence or absence of level crossings in the spectrum of a single MPS transfer matrix is a clear indicator of the occurrence of non-analytic cusps.

Conventional techniques for MPS time evolution in the thermodynamic limit \cite{ITEBD_supp} relying on an efficient Trotter-Suzuki decomposition \cite{Trotter_supp,Suzuki_supp} of the time-evolution operator $U(t)=\exp(-\rmi \mathcal{H} t)$ can be employed for long-range interacting systems, however only by introducing additional approximations \cite{Pollmann2015_supp}. MPS tangent-space methods based on the time-dependent variational principle (TDVP) \cite{TDVP_supp,MPS_tangent_supp, TDVP_Uni_supp} on the other hand directly integrate the time-dependent Schr\"odinger equation within the variational manifold of MPS and therefore only require the application of the Hamiltonian onto an MPS, thus avoiding an approximate Trotter-Suzuki decomposition of $U(t)$.

In order to efficiently apply a Hamiltonian with (e.g.~decaying power-law) long-range interactions onto an MPS, we expand it in a sum of exponentially decaying interaction terms, which we outline in \Sec{sec:LR}. We then describe the algorithm we use to simulate the real-time evolution of a pure state on a spin chain under such a Hamiltonian in the thermodynamic limit in \Sec{sec:algorithm}. In \Sec{sec:rf1} and \Sec{sec:rf2} we finally describe how to efficiently calculate the Loschmidt-echo return rate in the framework of iMPS and how non-analyticities in the form of cusps naturally arise due to level crossings in the (mixed) MPS transfer matrix arising in the overlap between the time-evolved and initial states.

% In this section we give a short description of the algorithm used to simulate the real-time evolution of a pure state on a spin chain under a Hamiltonian with long-range interactions in the thermodynamic limit. In our case, this Hamiltonian is the long-range transverse-field Ising model, but our algorithm applies to other general models.
% Moreover, as explained in detail in the following, DPT-II can be very straightforwardly detected from level crossings in the eigenvalues of the MPS transfer matrix.

\subsection{Long-range interactions}
\label{sec:LR}
We assume the Hamiltonian to contain a sum of two-body interactions of the form $h^{(2)}_{ij} = f(|j-i|)\,O_{i}O_{j}$, where operators $O_{i}$ act on a single site $i$ and commute when acting on different sites $[O_{i},O_{j}]=0,\,i\neq j$. Specifically, we consider the long-range transverse-field Ising model

\begin{align}\label{eq:LRTFIM}
\mathcal{H}=-J\sum_{j>i=1}^L\frac{\sigma^z_i\sigma^z_j}{|i-j|^{\alpha}}-h\sum_i\sigma^x_i,
\end{align}

\noindent where $\sigma^{x,z}_{i}$ are Pauli matrices acting on site $i$, $J>0$ is the spin-spin coupling constant, $h$ is the transverse magnetic field, and $L$ is the number of sites. We consider the thermodynamic limit $L\to\infty$.  The distance function decays as a power law $f(n)=n^{-\alpha}$ and we assume $f(n)$ to be well approximated \cite{Crosswhite2008} by a sum of $M$ exponentials, \textit{i.e.}~$f(n)\approx\sum_{k=1}^{M}c_{k} \lambda_{k}^{n-1}$, with $\lambda_{k}<1$ and $n>0$. With this approximation the overall Hamiltonian is then given by
\begin{equation}
%  H=\sum_{k=1}^{K}\sum_{j>i}c_{k} \lambda_{k}^{j-i-1}\, O_{i}O_{j}
 \mathcal{H}=-J\sum_{k}\sum_{j>i} c_{k} \lambda_{k}^{j-i-1}\, \sigma^{z}_{i}\sigma^{z}_{j} - h\sum_{i}\sigma^{x}_{i}.
 \label{eq:Ham}
\end{equation}
For an infinite system we perform a non-linear least-squares fit of $f(n)$ with a suitable number of exponentials over a distance $N$ large enough such that $f(N)<\varepsilon_{f}$, where $\varepsilon_{f}$ is of the order $\mathcal{O}(10^{-6})$ - $\mathcal{O}(10^{-8})$ and the largest residuals are of the order $\mathcal{O}(10^{-5})$. This usually amounts to $M$ in the range of $5-25$. For further details on this implementation, we refer the reader to Ref.~\onlinecite{Crosswhite2008}.

\subsection{MPS real-time evolution}
\label{sec:algorithm}
To simulate the real-time evolution of a pure quantum state $\ket{\psi}$ within the variational space of MPS with respect to \eqref{eq:Ham}, we adapt the algorithm of Ref.~\onlinecite{TDVP_Uni_supp} to the thermodynamic limit. The method of Ref.~\onlinecite{TDVP_Uni_supp} integrates the time-dependent Schr\"odinger equation $\partial_{t}\ket{\psi}=-\rmi\,\mathcal{H}\ket{\psi}$ by applying the time-dependent variational principle (TDVP) onto the variational manifold of MPS \cite{TDVP_supp}, but uses a Lie-Trotter splitting scheme of the projector onto the tangent space of the variational manifold to directly integrate the effective differential equations for the MPS tensors. Due to this splitting scheme it is necessary to evolve the state in small time steps $\tau$ only. Note, however, that this is conceptually very different from a Lie-Trotter splitting of the global time evolution operator $U(t)$, as employed in many conventional MPS time-evolution algorithms. For details on notation and the algorithm we refer the reader to Refs.~\onlinecite{TDVP_supp,TDVP_Uni_supp}.

In the following we describe the evolution of a translation-invariant MPS in the thermodynamic limit (iMPS) by a small time step $\tau$. We assume the state at time $t$ to be given in terms of a translation-invariant iMPS in the mixed canonical form \cite{Uli_supp}, \textit{i.e.}~the state is well approximated by an MPS given by (site-independent) MPS tensors $A_{L}^{s}$ and $A_{R}^{s}$ in the left and right gauge, and a bond matrix $C$ whose singular values are the Schmidt values of a bipartition of the state
\begin{equation}
 \ket{\psi} = \sum_{\bf{s}}(\ldots A_{L}^{s_{n-2}}A_{L}^{s_{n-1}}C A_{R}^{s_{n}} A_{R}^{s_{n+1}}\ldots)\ket{\bf{s}},
\end{equation} 
where $n$ is an arbitrary site on the chain. This also defines a center-tensor 
\begin{equation}
A_{C}^{s}=A_{L}^{s}C=CA_{R}^{s},
\label{eq:AC}
\end{equation} 
and the gauge conditions read
\begin{align}
 \sum_{s}{A_{L}^{s}}^{\dagger}A_{L}^{s} &= \unity &  \sum_{s}A_{L}^{s}CC^{\dagger}{A_{L}^{s}}^{\dagger} &= CC^{\dagger},\label{eq:Lgauge}\\
 \sum_{s}A_{R}^{s}{A_{R}^{s}}^{\dagger} &= \unity & \sum_{s}{A_{R}^{s}}^{\dagger}C^{\dagger}C\,A_{R}^{s} &= C^{\dagger}C.\label{eq:Rgauge}
\end{align}

\begin{figure}[tb]
 \centering
 \includegraphics[width=1.0\linewidth,keepaspectratio=true]{\figpath/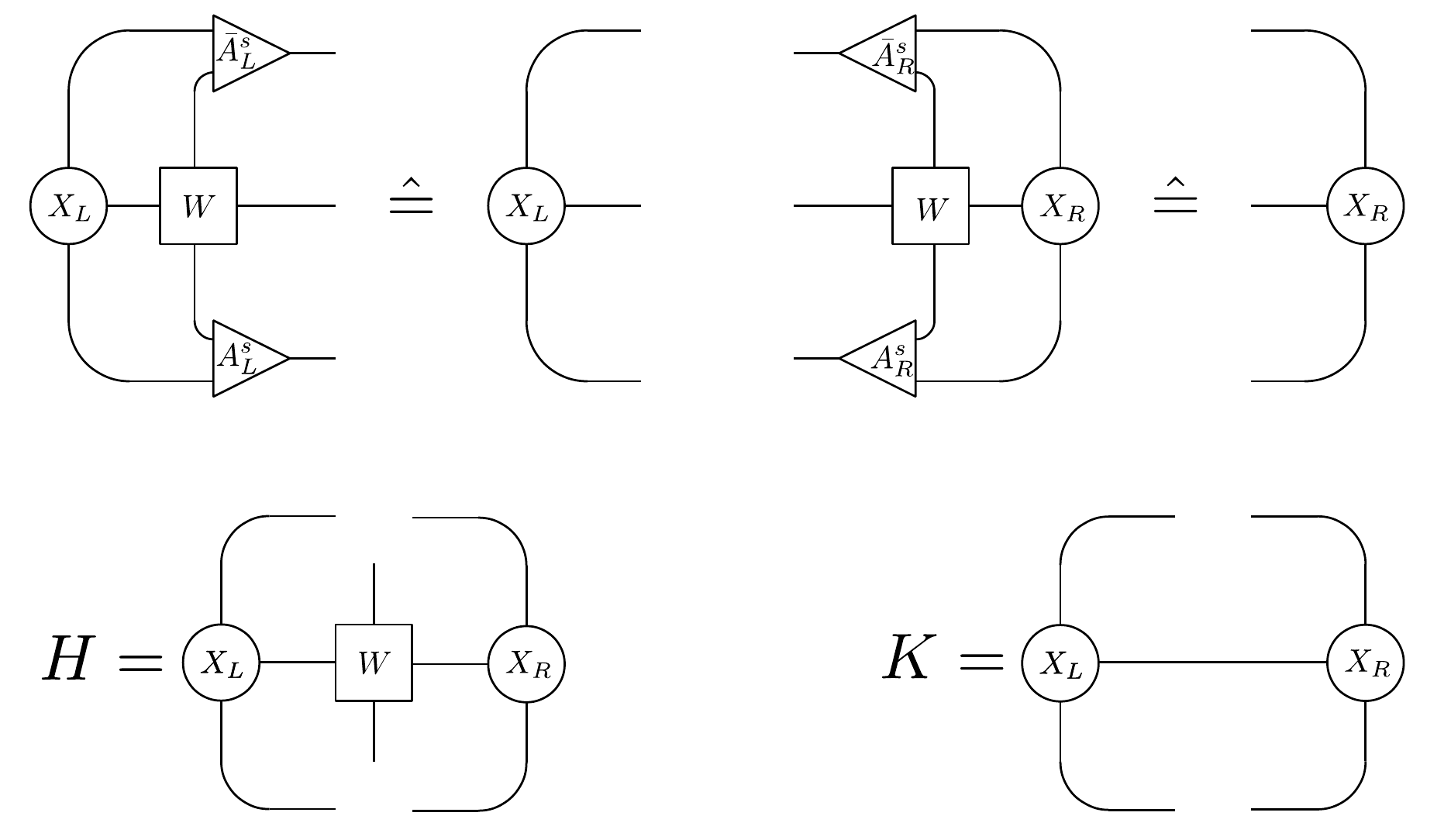}
 % Heff.pdf: 0x0 pixel, 0dpi, 0.00x0.00 cm, bb=
 \caption{Fixed point relations for the left and right MPO transfer matrices \eqref{eq:MPOTM} and definition of the effective Hamiltonians $H$ for $A_{C}^{s}$ and $K$ for $C$. The symbol $\hat{=}$ refers to equality up to terms contributing to the energy expectation value per site \cite{mpoinf_supp} (for more details see also Appendix C.2 in Ref.~\onlinecite{VUMPS2017}).}
%  contribution extensive in the systems size, corresponding to the (infinite) current energy expectation value of the global system Hamiltonian $\mathcal{H}$ \cite{mpoinf}.}
 \label{fig:Heff}
\end{figure}

\noindent We represent the Hamiltonian \eqref{eq:Ham} in terms of a Matrix Product Operator (MPO) \cite{mpo1_supp,mpo2_supp,mpo3_supp,mpolong_supp,Crosswhite2008} $W$ with bond dimension $d_{W}=K+2$. The projection of $\mathcal{H}\ket{\psi}$ onto the MPS tangent plane defines effective Hamiltonians $H$ for $A_{C}^{s}$ and $K$ for $C$ \cite{TDVP_Uni_supp}. In order to calculate these efficiently it is necessary to determine the left and right (quasi) fixed points of the MPO transfer matrices (c.f.~Ref.~\onlinecite{mpoinf_supp} and \Fig{fig:Heff})
\begin{equation}
 \mathcal{T}_{L/R}^{ab}=\sum_{ss'} W^{ab}_{s's}\bar{A}_{L/R}^{s'}\otimes A_{L/R}^{s},
 \label{eq:MPOTM}
\end{equation} 
where a bar denotes complex conjugation and indices $a,b\in[1,\ldots,d_{W}]$ (see \Fig{fig:Heff}).

The algorithm in Ref.~\onlinecite{TDVP_Uni_supp} for a \textit{finite} chain proceeds by performing the following steps sequentially for each site $n$ of the lattice (e.g.~in a left-to-right sweep)
\begin{enumerate}
 \item Calculate $H(n)$ as a function of $\tilde{A}_{L}^{s}(m<n)$ and $A_{R}^{s}(m>n)$.
%  \item forming $A_{C}^{s}(n) = C(n-1)A_{R}^{s}(n)$ and 
 \item Evolve $\tilde{\mathbf{A}}_{C}(n) = \exp(-\rmi\tau\,H(n))\,\mathbf{A}_{C}(n)$ forward in time.
 \item Split $\tilde{A}_{C}^{s}(n)=\tilde{A}_{L}^{s}(n)\,\tilde{C}(n)$.
 \item Calculate $K(n)$ as a function of $\tilde{A}_{L}^{s}(m\leq n)$ and $A_{R}^{s}(m>n)$.
 \item Evolve  $\mathbf{C}(n)=\exp(+\rmi\tau\,K(n))\,\tilde{\mathbf{C}}(n)$ \textit{backward} in time and form $A_{C}^{s}(n+1) = C(n)\,A_{R}^{s}(n+1)$.
\end{enumerate}
Here $\tilde{A}_{L}^{s}(n)$ refers to tensors updated in previous steps and bold symbols refer to vectorizations of the tensors. Note that both $H$ and $K$ are calculated from \textit{evolved} $\tilde{A}_{L}^{s}$ and as-of-yet-unevolved $A_{R}^{s}$. In the thermodynamic limit, $A_{L}^{s}$ and $A_{R}^{s}$ correspond to the same state and we want to evolve both of them at the same time. The adapted algorithm given in Table~\ref{tab:algo} achieves this.

The quantities $\varepsilon_{L}$ and $\varepsilon_{R}$ can be used as a measure of the quality of a time step and of the constraint $A_{L}^{s}C=CA_{R}^{s}$ \eqref{eq:AC}. We reorthonormalize the state according to \eqref{eq:Lgauge} and \eqref{eq:Rgauge} if $\varepsilon_{L}$ and $\varepsilon_{R}$ rise above a certain threshold $\varepsilon_G$ \cite{footnoteA}. This new algorithm corresponds to a first-order splitting scheme \cite{footnote1_supp} with an error scaling of $\mathcal{O}(\tau^{2})$ \cite{TDVP_Uni_supp}. 

Notice also, that just as time-dependent density matrix renormalization group (DMRG) \cite{tDMRG1_supp, tDMRG2_supp} replaces the effective eigenvalue problem step of ground-state DMRG \cite{DMRG1_supp, DMRG2_supp} with a time evolution step, the above scheme replaces the effective eigenvalue problem step of the new ground state algorithm (VUMPS) presented in Ref. \onlinecite{VUMPS2017} with a time evolution step.

We increase the bond dimension of the MPS whenever the smallest of the Schmidt values (which are the singular values of $C$) rises above a certain threshold $\varepsilon_S$. For this we use the procedure presented in Appendix B of Ref.~\onlinecite{VUMPS2017}. In practice we use a time step of $\tau=0.002$, a reorthonormalization threshold of $\varepsilon_G = 10^{-8}$, a bond dimension increase threshold of $\varepsilon_S=10^{-9}$ and a maximum bond dimension of $D_{\rm max}\approx 350$, which -- depending on $\mathcal{H}$ -- limits the maximum reachable simulation time $t_{\rm max}$ severely due to the (at worst) linear increase of entanglement entropy \cite{blowS_supp,Prosen07_supp}.

\begin{table}[t]
\begin{enumerate}
 \item Calculate $H$ and $K$ from current $A_{L}^{s}$ and $A_{R}^{s}$ and form $A_{C}^{s}=A_{L}^{s}C=CA_{R}^{s}$.
 \item Evolve $\tilde{\mathbf{A}}_{C} = \exp(-\rmi\tau\,H)\,\mathbf{A}_{C}$ forward in time.
 \item Evolve $\tilde{\mathbf{C}} = \exp(-\rmi\tau\,K)\,\mathbf{C}$ \textit{forward} in time.
 \item Determine the optimal updated $\tilde{A}_{L}^{s}$ and $\tilde{A}_{R}^{s}$ by minimizing $\varepsilon_{L}=\lVert\sum_{s}\tilde{A}_{C}^{s} - \tilde{A}_{L}^{s}\tilde{C}\rVert_{2}$  and $\varepsilon_{R}=\lVert\sum_{s}\tilde{A}_{C}^{s} - \tilde{C}\tilde{A}_{R}^{s}\rVert_{2}$ under the constraints $\sum_{s} \tilde{A}_{L}^{s \dagger} \tilde{A}_{L}^{s} = \sum_{s} \tilde{A}_{R}^{s} \tilde{A}_{R}^{s \dagger}  = \unity$.
 \item (optional) If $\varepsilon_{L}$ or $\varepsilon_{R}$ rise above some threshold $\varepsilon_{G}$, reorthonormalize the state.
\end{enumerate}
\caption{One time step of the novel time evolution algorithm suitable for simulating systems with long-range interactions. It is obtained as the thermodynamic limit generalization of the finite system algorithm of Ref. \cite{TDVP_Uni_supp}. It also improves upon the original TDVP algorithm \cite{TDVP_supp} by not relying on taking possibly ill-defined inverses.}
\label{tab:algo}
\end{table} 
% \clearpage
\subsection{Loschmidt echo and the return rate}
\label{sec:rf1}

With the definition of the Loschmidt amplitude
\begin{equation}
G(t)=\braket{\psi(0)|\psi(t)}=\braket{\psi(0)|e^{-\rmi \mathcal{H} t}|\psi(0)},
\label{eq:L_echo}
\end{equation}
the return probability rate function per site
\begin{align}
r(t)=-\lim_{L\to\infty}\frac{1}{L}\log|G(t)|^2
\label{eq:ratefun}
\end{align}
corresponds to (minus) the logarithm of the dominant eigenvalue of the mixed MPS transfer matrix \cite{Karrasch13_supp}
\begin{equation}
 \mathcal{T}(t)=\sum_{s}\bar{A}^{s}(0)\otimes A^{s}(t)
 \label{eq:MTM}
\end{equation} 
between MPS tensors at time zero and $t$, whose spectral radius is $\rho(\mathcal{T}(t))\leq1$ \cite{footnote2_supp}. If $\epsilon_{i}(t)$ are the eigenvalues of \eqref{eq:MTM} in descending order by magnitude (\textit{i.e.}~$\epsilon_{1}(t)$ being the largest), then we define the rate-function branches
\begin{equation}
 r_{i}(t) = -2\log|\epsilon_{i}(t)|,
 \label{eq:ratefun_branches}
\end{equation} 
and the rate function \eqref{eq:ratefun} is simply $r(t)=r_{1}(t)$, with all other $r_{i>1}>r_{1}(t)$.

\begin{figure}[t]
 \centering
 \includegraphics[width=1.0\linewidth,keepaspectratio=true]{\figpath/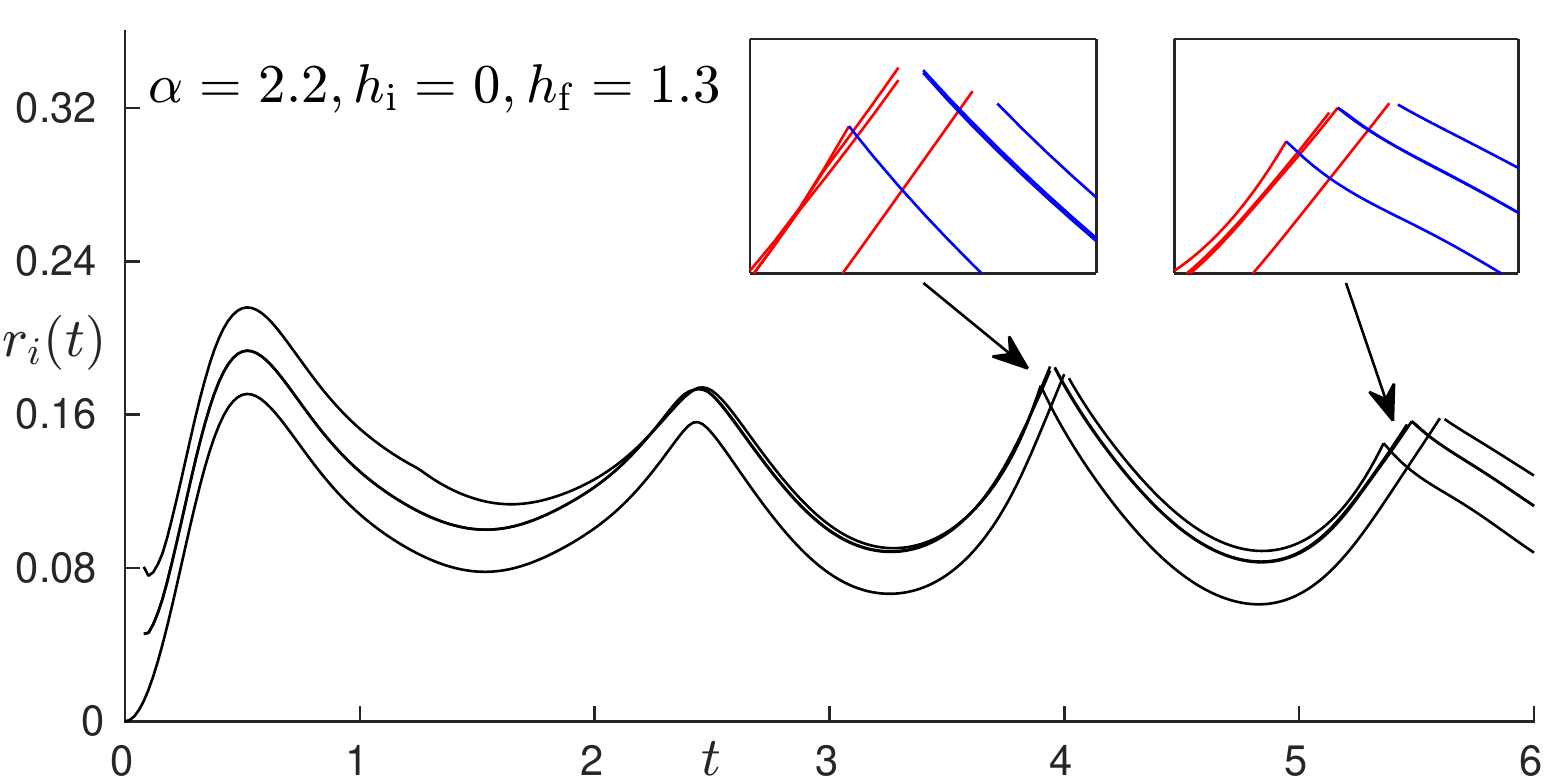}
 % Plot8_crossing_example.pdf: 0x0 pixel, 0dpi, 0.00x0.00 cm, bb=
 \caption{Example of how cusps in the return rate $r(t)$ can be detected through level crossings of the rate-function branches $r_{i}(t)$ \eqref{eq:ratefun_branches}. Shown are the first four $r_{i}(t)$ for a quench from $h_{\rm i}=0$ to $h_{\rm f} =1.3$ and $\alpha=2.2$, where the \textit{lowest} branch represents the rate function \eqref{eq:ratefun}. At $t\approx2.4$ there are no level crossings, $r(t)$ is thus smooth. At times $t\approx 4$ and $t\approx 5.5$ there are level crossings in the rate-function branches, which cause cusps in $r(t)$. The example shows anomalous cusps, however regular cusps (or in fact any other type thereof) show up due to the same mechanism. The insets show a magnification of these cusps, where branches that are going up are colored red and branches coming down are colored blue for better visualization.
 }
 \label{fig:crossing_ex}
\end{figure}

Nonanalyticities of \eqref{eq:ratefun} in $t$ in the form of cusps arise due to level crossings in the eigenspectrum of \eqref{eq:MTM}, which is a feature characteristic of first-order phase transitions. The occurrence of such non-analyticities can thus nicely be anticipated by calculating and following the first few rate-function branches. Cusps arise at branch crossings, where a higher branch becomes the lowest one at some time $t_{c}$ (see \Fig{fig:crossing_ex}).

The mixed MPS transfer matrix \eqref{eq:MTM} represents an approximation of the true Quantum transfer matrix (QTM) of a real-time path-integral formulation of \eqref{eq:L_echo}, which can be understood as the result of a renormalization procedure of an effective impurity problem for the true QTM along the real time axis \cite{Zauner15_supp, Rams2015_supp, Bal2015_supp}. Due to this fact, the eigenvalues of the QTM which are reproduced with the highest accuracy are the dominant ones, with smaller and smaller ones being reproduced with less and less accuracy. This also means that the lowest few rate function branches \eqref{eq:ratefun_branches} are the most accurate ones with a given finite bond dimension, and the approximation of the return rate function by $r_{1}(t)$ is valid.

\begin{figure*}
 \centering
 \includegraphics[width=0.8\linewidth,keepaspectratio=true]{\figpath/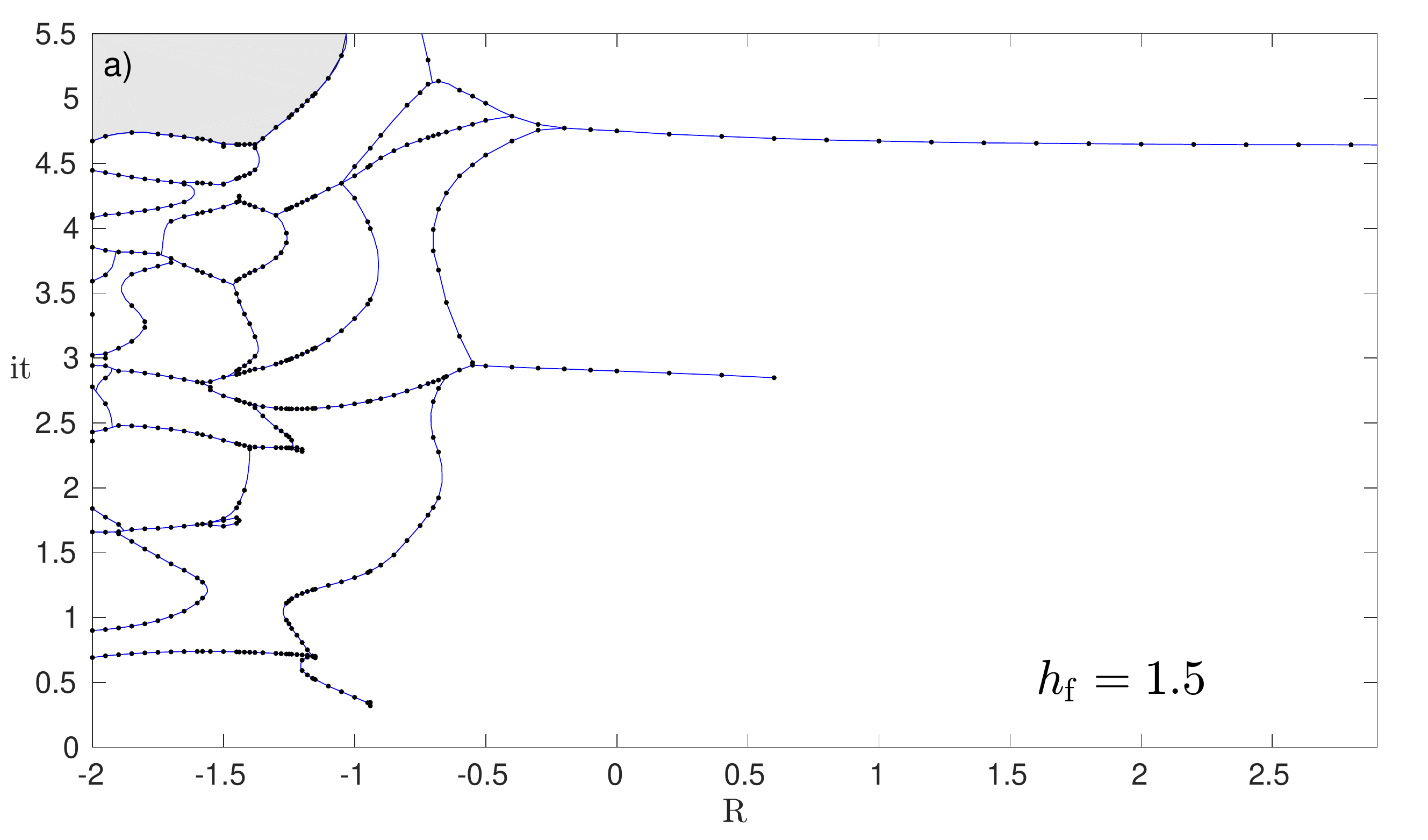}
 \includegraphics[width=0.8\linewidth,keepaspectratio=true]{\figpath/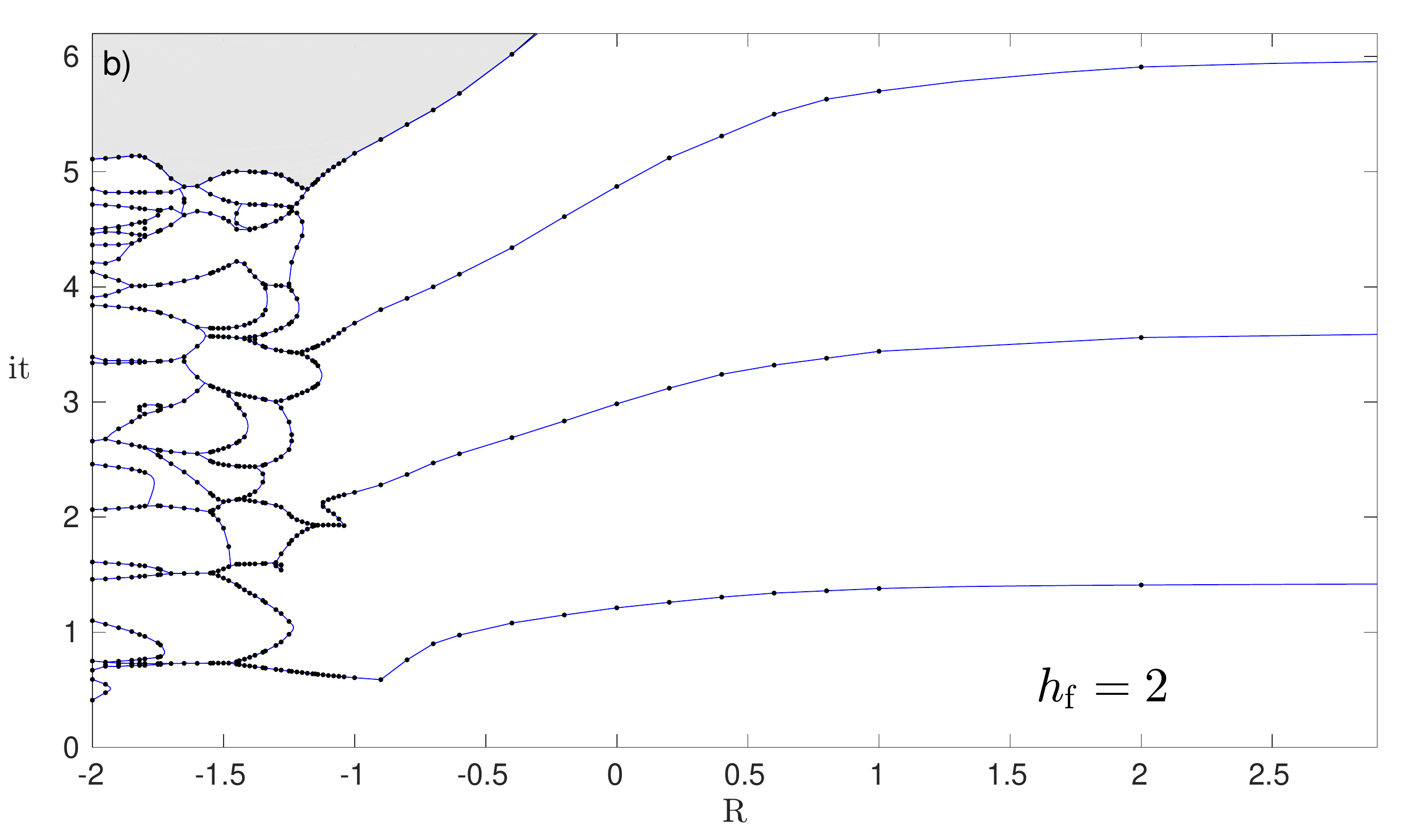}
 % FZ_a2.2_hf1.5: 0x0 pixel, 300dpi, 0.00x0.00 cm, bb=
 \caption{Fisher-zero lines for $\alpha=2.2$ and $h_{\rm i}=0$ in the complex $z=R+\rmi t$ plane. We show data for quenches to (a) $h_{\rm f}=1.5<h^{\rm II}_{\rm c,z}$ in the anomalous phase and (b) $h_{\rm f}=2>h^{\rm II}_{\rm c,z}$ in the regular phase. It can be seen that in (a) the lowest FZL cutting the imaginary axis ($R=0$) terminates around $R\approx0.6$ (this process is depicted in \Fig{fig:FZL_end}). With decreasing $h_{\rm f}$, this endpoint (and the endpoints of all higher FZL cutting the imaginary axis) move to smaller and smaller $R$, until the endpoints cross the imaginary axis and the cusps in $r(t)$ vanish one by one, starting at early times $t$. The black dots represent data points, the blue lines are guides to the eye. Numerical data was not sufficient for display in the top left greyed-out areas in both (a) and (b).
 }
 \label{fig:FZL}
\end{figure*}

\begin{figure*}
 \centering
 \includegraphics[width=\linewidth,keepaspectratio=true]{\figpath/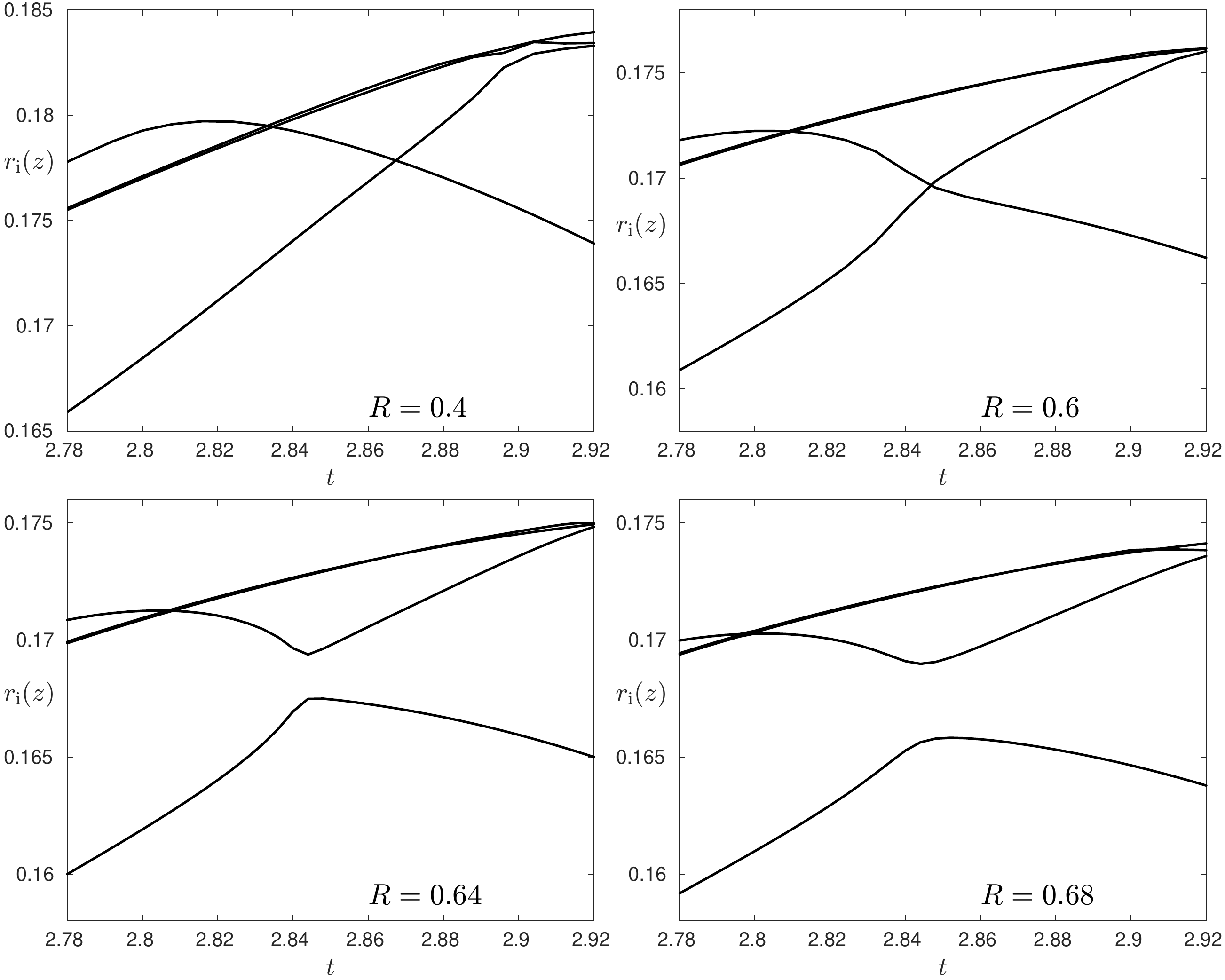}
 % FZ_a2.2_hf1.5: 0x0 pixel, 300dpi, 0.00x0.00 cm, bb=
 \caption{Plots of the rate function branches $r_{\rm i}(z)$ with complex argument $z=R+\rmi t$ vs.~real time $t$ for four different (fixed) values of $R=\{0.4, 0.6, 0.64,0.68\}$, showing the mechanism responsible for the termination of FZL in the complex plane in the anomalous phase (here in particular for the lowest FZL in \Fig{fig:FZL}). It can be seen that the crossing of the two lowest branches present in the top two panels has disappeared with increasing $R$ in the bottom two panels. Notice that this is not due to one branch moving up such that they no longer cross, but by a disconnecting process that roughly leaves the branches in their general position, but turns the crossing more and more into an avoided crossing. This feature is characteristic for the disappearance of anomalous cusps, whereas regular cusps always vanish due to relative movement of the branches only.
 }
 \label{fig:FZL_end}
\end{figure*}

\subsection{Alternative calculation of the rate function}
\label{sec:rf2}

Trivially, \eqref{eq:L_echo} can also be written as
\begin{align}
\braket{\psi(0)|\psi(t)} &=   \braket{\psi(0)|\exp(-i\mathcal{H} t)|\psi(0)} \notag\\ 
&=\left(\bra{\psi(0)}\exp(-i\mathcal{H} t/2)\right)\left(\exp(-i\mathcal{H} t/2)\ket{\psi(0)}\right)\notag \\
&= \braket{\psi(-t/2)|\psi(t/2)},
\end{align} 
where in the present case the backward-evolved state $\ket{\psi(-t/2)}$ can be obtained from the forward-evolved state as $\ket{\psi(-t/2)}=\overline{\ket{\psi(t/2)}}$, \textit{i.e.}~by complex conjugation. We exploit this fact to evaluate the rate function at time $t$ with the MPS at time $t/2$, \textit{i.e.}~we use the leading eigenvalues of 
\begin{equation}
  \tilde{\mathcal{T}}(t)=\sum_{s}A^{s}(t/2)\otimes A^{s}(t/2).
 \label{eq:MTM2}
\end{equation} 
Thus, $r(t)$ can be calculated both from \eqref{eq:MTM} at time $t$ and also from \eqref{eq:MTM2} at time $t/2$, and the agreement of both serves as an additional check for the validity of the numerical results. Return rates of the LR-TFIM computed in the framework of iMPS can be found in Ref.~\onlinecite{Halimeh2016b}.

\section{Fisher-Zero Lines}
\label{sec:FZ}

In this section we present additional evidence for the qualitative difference between regular and anomalous cusps by calculating Fisher-zero lines (FZL) in the complex planes for various quenches from $h_{\rmi}=0$ for $\alpha=2.2$. In particular we show data for a quench to $h_{\rm f}=1.5<h^{\rm II}_{\rm c,z}$ (showing anomalous cusps), and a quench to $h_{\rm f}=2>h^{\rm II}_{\rm c,z}$ (showing regular cusps). The plots in \Fig{fig:FZL} show that the FZL crossing the imaginary axis are also qualitatively different for the two different quenches, which further corroborates that the two types of cusps are distinct.

We consider the boundary partition function
\begin{equation}
\mathcal{Z}_{b}(z)=\braket{\psi(0)|e^{-z \mathcal{H}}|\psi(0)}
\label{eq:bdry_Z_supp}
\end{equation} 
in the complex plane $z=R + \rmi t$, with $R,t\in\mathbb{R}$.
The complex zeros $z_{j}$ of $\mathcal{Z}_{b}(z)$ cause the free energy density analog
\begin{equation}
 f(z) = -\frac{1}{L}\log |\mathcal{Z}_{b}(z)|^{2}
\end{equation} 
to be non-analytic at these points. In the limit of infinite system size $L\to\infty$ these zeros tend to arrange along lines, a fact which was pointed out by Fisher who studied the analytic properties of the thermal partition function of the classical two-dimensional Ising model in the complex temperature plane \cite{Fisher1965}. As already discussed, this is similar to the famous Lee-Yang circle theorem \cite{Yang1952,Lee1952}, which states that the zeros of a classical thermal partition function $\mathcal{Z}_{\rm th}(\beta,\mu)=\exp[-\beta(H-\mu N)]$ form circles in the complex \textit{fugacity}-plane, where $z=\exp(\beta \mu)$ is the fugacity, with $\beta$ real and $\mu$ complex and $N$ the particle number.

% In fact, it had been shown by Lee and Yang earlier, \cite{LeeYang1_supp,LeeYang2_supp}, that the zeros of the thermal partition function of the classical two-dimensional Ising model form circles in the complex \textit{magnetic field} plane, however no such rigorous ...

% Unlike the famous Lee-Yang circle theorem , which states that the zeros of the thermal partition function of the classical two-dimensional Ising model in the complex \textit{magnetic field} plane form circles

Nonanalyticities in the return rate function $r(t)$ therefore appear whenever an FZL crosses the imaginary axis $R=0$, such that $z=\rmi t$ and \eqref{eq:bdry_Z_supp} turns into the Loschmidt amplitude $G(t)$, and equivalently, $f(z)$ turns into the return rate function $r(t)$.

Numerically, we can calculate the FZL by utilizing
\begin{align}\nonumber
 \mathcal{Z}_{b}(z) &= \braket{\psi(0)|e^{-(R+\rmi t) \mathcal{H}}|\psi(0)}\\\nonumber
 &= \braket{\psi(0)|e^{-R/2\;\mathcal{H}} e^{-\rmi t \mathcal{H}} e^{-R/2\;\mathcal{H}}|\psi(0)}\\
 &= \braket{\psi_{R/2}(0)|e^{-\rmi t \mathcal{H}}|\psi_{R/2}(0)},
\end{align}
where we have defined $\ket{\psi_{R/2}(0)}=e^{-R/2\;\mathcal{H}}\ket{\psi(0)}$ in the last line. To scan the complex plane for non-analyticities of $f(z)$, we therefore prepare the system in state $\ket{\psi_{\tilde{R}/2}(0)}$ for some fixed $\tilde{R}$ (e.g.~through imaginary-time evolution) and then perform real-time evolution. A non-analyticity at $t=\tilde{t}$ during this evolution thus implies a Fisher zero at $\tilde{z}=\tilde{R}+\rmi\tilde{t}$. Note that a similar strategy for calculating FZL in the NN-XXZ model has also been employed in Ref.~\onlinecite{Andraschko2014}.

In \Fig{fig:FZL} we show examples for $\alpha=2.2$ and $h_{\rm_i}=0$, with panel (a) for $h_{\rm f}=1.5<h^{\rm II}_{\rm c,z}$ in the anomalous phase where $r(t)$ shows \textit{anomalous} cusps for regular real-time evolution, and panel (b) for $h_{\rm f}=2>h^{\rm II}_{\rm c,z}$ in the regular phase, where $r(t)$ shows \textit{regular} cusps. In both cases the Fisher zeros form intricate patterns in the $R<0$ half plane, reminiscent of quenches within the same phase in the NN-TFIM model (cf.~e.g.~Fig.~1 in Ref.~\onlinecite{Heyl2013}) or more complicated quenches in the NN-XXZ model \cite{Andraschko2014}.

In both cases, a number of FZL emanate from the $R<0$ half plane to cross the imaginary axis and cause cusps in the rate function $r(t)$ in the case of pure real-time evolution ($R=0$). In the regular phase these FZL look very similar to the case of the NN-TFIM, \textit{i.e.}~they continue on to $R\to\infty$ while monotonically increasing (in $t$ vs.~$R$, see \Fig{fig:FZL}(b)). In contrast, in the anomalous phase, these emanating FZL terminate at some finite $R$ (this can be seen in \Fig{fig:FZL}(a) for $h_{\rm f}=1.5$ for the lowest FZL crossing the imaginary axis, which terminates at $R\approx 0.6$). Additionally, the emanating FZL for $h_{\rm f}<h^{\rm II}_{\rm c,z}$ do not increase monotonically (at least within the considered simulation parameters), which necessarily requires them to terminate at some finite $R$, as there can be no non-analyticities in $f(z)$ for $R\to\infty$ \cite{footnoteB}. Upon decreasing $h_{\rm f}$ the endpoints of these lines move to smaller and smaller $R$. Whenever such an endpoint crosses the imaginary axis, the corresponding anomalous cusp consequently vanishes in the rate function for real-time evolution. While further decreasing $h_{\rm f}$ more and more anomalous cusps thus vanish one by one at larger and larger times. This observation is also consistent with Ref.~\onlinecite{Halimeh2016b} (cf.~Fig. 7 therein).

\begin{figure}[htb]
 \centering
 \includegraphics[width=1.0\linewidth,keepaspectratio=true]{\figpath/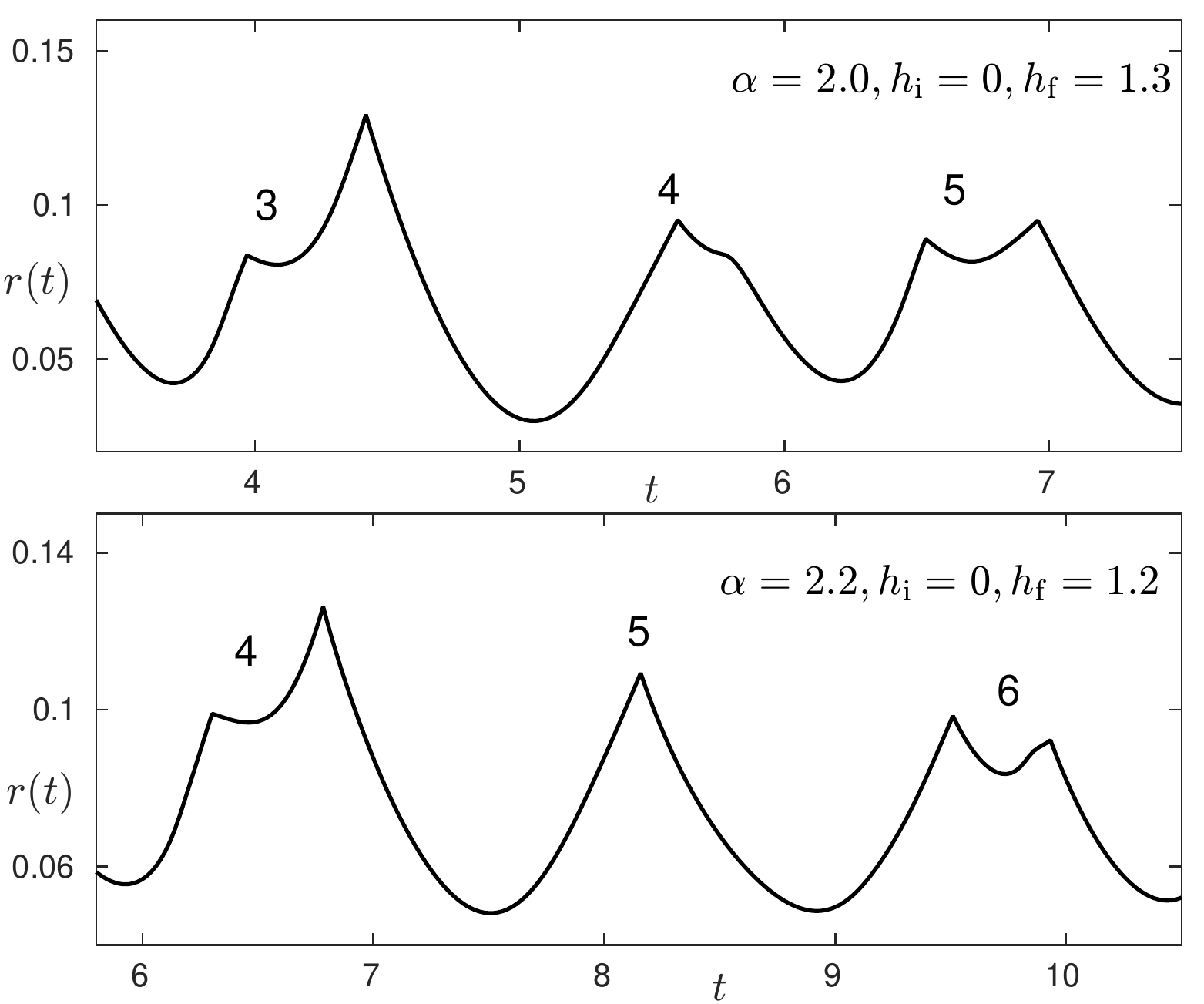}
 % Plot8_crossing_example.pdf: 0x0 pixel, 0dpi, 0.00x0.00 cm, bb=
 \caption{Examples of double cusps for quenches within the anomalous phase. Here for $\alpha=2$, the 3rd and 5th cusps are double cusps, for $\alpha=2.2$ the 4th and 6th cusps are double cusps.
 }
 \label{fig:double_cusps}
\end{figure}

Cusps in the regular phase appear or disappear solely due to individual rate function branches moving relative to one another upon varying $h_{\rm f}$. Anomalous cusps on the other hand can also disappear due to branch crossings turning into \textit{avoided} crossings. Such a scenario is depicted in \Fig{fig:FZL_end}, which demonstrates how the lowest FZL cutting the imaginary axis in \Fig{fig:FZL} terminates around $R\approx0.6$.

These additional qualitative differences between FZL in the regular and the anomalous phase further corroborate the existence of two distinct dynamical critical phases for quenches above or below $h^{\rm II}_{\rm c,z}$ and $\alpha\lesssim2.3$.
% For $z=\rmi t$, this is exactly the overlap between the time evolved state and its initial self. 

\begin{figure}[t]
 \centering
 \includegraphics[width=1.0\linewidth,keepaspectratio=true]{\figpath/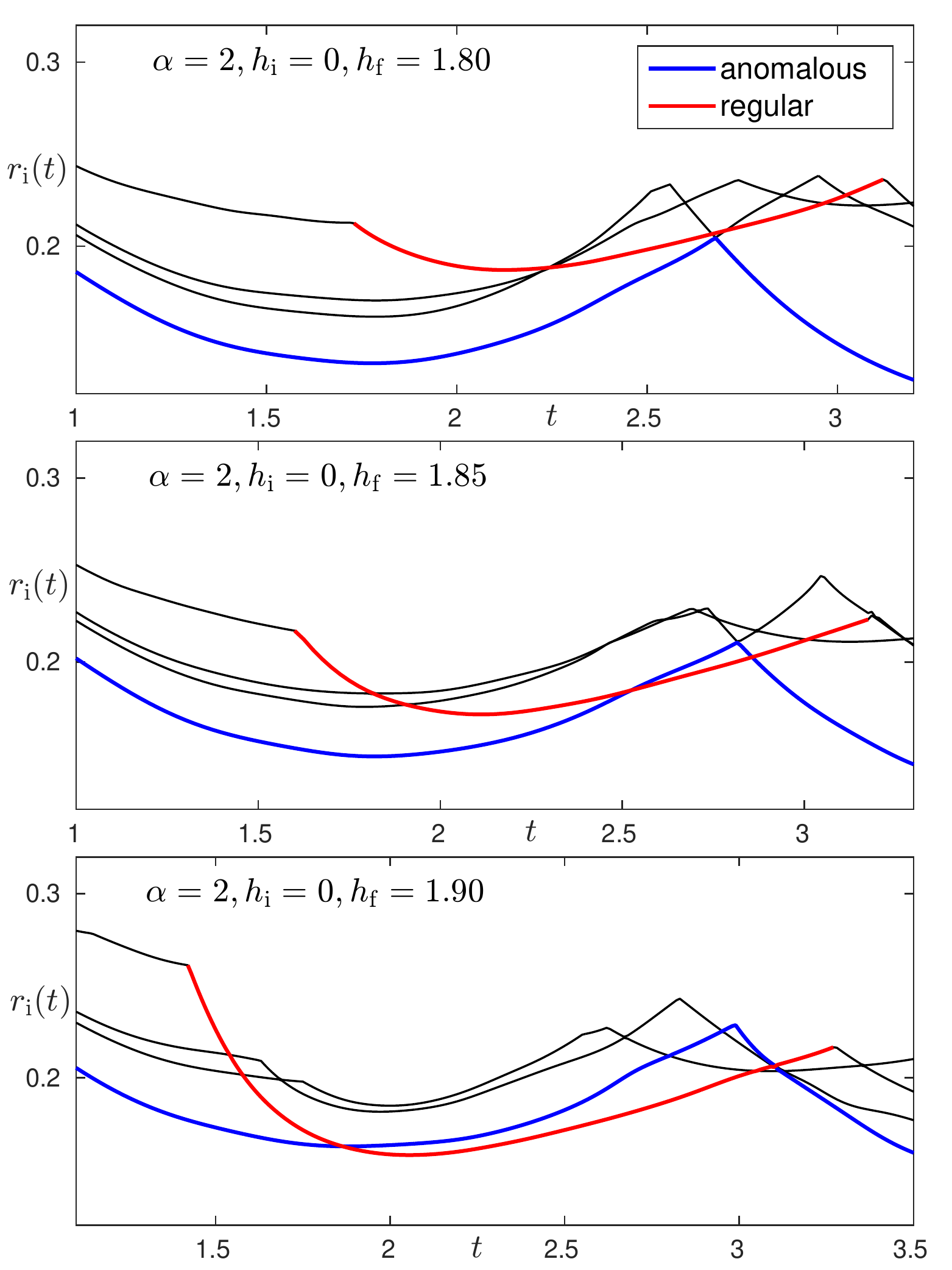}
 % Plot8_crossing_example.pdf: 0x0 pixel, 0dpi, 0.00x0.00 cm, bb=
 \caption{Example of rate-function branches $r_{i}(t)$ responsible for developing regular cusps (red) coming down and taking over branches responsible for developing anomalous cusps (blue) with increasing $h_{\rm f}$ (here $\hczII=1.85$).
 }
 \label{fig:crossover}
\end{figure}

\section{Additional anomalous behavior}
\label{sec:appendix}

\subsection{Double cusps}
\label{sec:double}
Some of the anomalous cusps that show up for quenches with $\alpha\lesssim2.3$ and $h_{\rm f}<h^{\rm II}_{\rm c,z}$ develop a double cusp structure in the sense that the tip of these cusps seems to be cut off by another rate-function branch coming down at exactly that time. Examples for such double cusps are shown in \Fig{fig:double_cusps}. The location of these double cusps also seems to drift with $\alpha$ (see also Fig.~7 in Ref. \onlinecite{Halimeh2016b}). The nature and origin of these double cusps remains elusive, although we suspect they could be due to FZL cutting the imaginary axis coming from the $R<0$ half-plane and returning again immediately. As such double cusps do not appear in the regular phase, they constitute another feature distinguishing both phases. It is however worth noting that such double cusps have previously also been observed for the transverse axial next-nearest-neighbor and tilted-field Ising model \cite{Karrasch13_supp}. We leave the investigation of this double-cusp feature open for future work, as it currently challenges the limits of our numerical capabilities.

\subsection{Crossover between regular and anomalous cusps}
\label{sec:crossover}

In this section we describe for quenches with $\alpha\lesssim 2.3$ and $h_{\rm i}=0$, how regular cusps replace anomalous cusps with $h_{\rm f}$ approaching and going over $h^{\rm II}_{\rm c,z}$. This process is in fact due to a group of rate-function branches $r_{i}(t)$ \eqref{eq:ratefun_branches} that are initially high up for $h_{\rm f}<h^{\rm II}_{\rm c,z}$, but come down with increasing $h_{\rm f}$, and -- after a narrow regime of coexistence around $\hczII$ -- eventually become lower than the branches responsible for developing  anomalous cusps. This new group of rate-function branches then develops regular cusps by crossing one another for all $h_{\rm f}>h^{\rm II}_{\rm c,z}$. Such a crossover situation is depicted in \Fig{fig:crossover} for $\alpha=2$. Further increasing $h_{\rm f}$ will cause all anomalous cusps to be completely covered by these new regular branches. The anomalous cusps could in principle still be followed and identified in the now higher-up anomalous rate-function branches, but they no longer contribute to the actual return rate function $r(t)$ (which is strictly given by the lowest branch) and cusps therein. Indeed, this provides further evidence of how the anomalous cusps are distinct from their regular counterparts in the sense that they do not simply morph into the regular cusps upon increasing $h_{\rm f}$ above the dynamical critical point $h^{\rm II}_{\rm c,z}$.

\section{Conclusion}
\label{sec:conclusion}
We have presented and discussed in detail how the Loschmidt-echo return rate (and non-analytic cusps therein, related to dynamical phase transitions) for a global quench scenario in the one-dimensional long-range (power-law interacting) transverse-field Ising model can be efficiently calculated in the framework of infinite Matrix Product States. We demonstrate in particular, that -- in contrast to finite-size approaches -- working directly in the thermodynamic limit enables an unambiguous identification of non-analyticities, which arise from level crossings in the eigenvalues of the Matrix Product State transfer matrix. We have further illustrated how the anomalous and regular dynamical critical phases, first reported in Ref.~\onlinecite{Halimeh2016b}, are distinct by showing that they arise due to qualitatively different Fisher-zero-line behavior. Whereas the regular phase constitutes cusps that arise due to the relative movement of so-called rate function branches of the Matrix Product State transfer matrix, the anomalous cusps are related to avoided crossings of these branches turning into actual crossings. Moreover, we have presented results showing double-cusp properties that further set the anomalous phase apart from its regular counterpart. Even though we cannot ascertain the origin of these double cusps, we do hypothesize that they arise due to Fisher-zero lines cutting the imaginary axis and then quickly turning around and cutting it again. Our work further validates the existence of the anomalous phase in the one-dimensional transverse-field Ising model for quenches from the $\mathbb{Z}_2$-symmetry-broken equilibrium phase to a final value of the transverse field below the dynamical critical point. Moreover, the method discussed in this work provides a versatile tool that is suitable to investigate, naturally in the thermodynamic limit, dynamical phase transitions in the context of non-analyticities in the Loschmidt-echo return rate, and properly characterize its different phases in various models, even though here we only tackle the transverse-field Ising chain with power-law interactions.

\section*{Acknowledgments}
We thank Damian Draxler, Jutho Haegeman, Frank Verstraete, Ian P. McCulloch, and Johannes Lang for inspiring and helpful discussions. V.~Z.-S. gratefully acknowledges support from the Austrian Science Fund (FWF): F4104 SFB
ViCoM and F4014 SFB FoQuS. The computational results presented have been achieved in part using the Vienna Scientific Cluster (VSC).

%----------------------------------------------
%\begin{thebibliography}{99}
%*********************************************************************************************

%--------------------------------------------------------

\begin{thebibliography}{9}
\bibitem{footnote1_supp}
Higher order schemes can be constructed in a slightly different way, but are computationally much more unfavorable than performing a first order step several times for a smaller time step.
\bibitem{footnote2_supp}
The gauge choices for $A^{s}(0)$ and $A^{s}(t)$ do not affect the eigenvalues of $\mathcal{T}(t)$ and are therefore irrelevant.
\bibitem{footnoteB}
For $R\to\infty$ the initial state is the ground state of the evolution Hamiltonian and therefore a stationary state of the evolution.
\end{thebibliography}
\end{document}